\documentclass[aps,prd,twocolumn,superscriptaddress,showpacs]{revtex4}
\usepackage{natbib,hyperref,ifthen} 
\usepackage{graphicx}
\usepackage{dcolumn}
\usepackage{bm}
\usepackage{natbib}
\usepackage{multirow}
\usepackage{epsfig}
\usepackage{amsmath}

\newcommand{\beq}{\begin{equation}}
\newcommand{\eeq}{\end{equation}}
\newcommand{\beqa}{\begin{eqnarray}}
\newcommand{\eeqa}{\end{eqnarray}}

\newcommand{\be}{\begin{equation}}
\newcommand{\ee}{\end{equation}}

\newcommand{\rmd}{{\rm d}}
\renewcommand{\vec}[1]{{\bf #1}}

\renewcommand{\vec}[1]{{\bf #1}}

\begin{document}

\title{Alignment of galaxy spins in the vicinity of voids}

\author{An\v{z}e Slosar} 
\affiliation{Berkeley Center for Cosmological
  Physics, Physics Department and Lawrence Berkeley National
  Laboratory,University of California, Berkeley California 94720, USA}
\affiliation{Faculty of Mathematics and Physics, University of
  Ljubljana, Slovenia}
\author{Martin White} 
\affiliation{Departments of Physics and Astronomy,
University of California, Berkeley, California 94720, USA}

\date{\today}

\begin{abstract}
  We provide limits on the alignment of galaxy orientations with the
  direction to the void center for galaxies lying near the edges of
  voids.  We locate spherical voids in volume limited samples of
  galaxies from the Sloan Digital Sky Survey using the HB inspired
  void finder and investigate the orientation of (color selected)
  spiral galaxies that are nearly edge-on or face-on. In contrast with
  previous literature, we find no statistical evidence for departure
  from random orientations.  Expressed in terms of the parameter $c$,
  introduced by Lee \& Pen to describe the strength of such an
  alignment, we find that $c<0.11(0.13)$ at 95\% (99.7\%) confidence
  limit within a context of a toy model that assumes a perfectly
  spherical voids with sharp boundaries.
\end{abstract}


\maketitle

\setcounter{footnote}{0}

\section{Introduction}
\label{sec:introduction}

Within the currently accepted paradigm for structure formation,
galaxies form through a complex series of mergers and accretion events
in a beaded filamentary network of dark matter halos and subhalos.
The cooling and condensation of gas within these dark matter hosts
leads to the visible signatures by which we recognize galaxies
\cite{1977MNRAS.179..541R,1977ApJ...211..638S,1978MNRAS.183..341W,1980MNRAS.193..189F,1984Natur.311..517B}.
While this broad picture has received impressive support in recent
years, many of the detailed processes remain poorly understood.  In
particular the manner in which feedback alters the rate at which gas
cools and the coupling between the angular momentum of the gas and the
dark matter -- both of which effect the sizes and shapes of disks --
are outstanding problems in galaxy formation.  We still cannot form
realistic looking disk galaxies from an {\it ab initio\/} simulation
in a cosmological context.

This lack is important not just for our understanding of galaxy
formation, but also because it impacts upon one of the most important
probes of structure formation and cosmology: weak gravitational
lensing (see \cite{2006glsw.book..269S} and references therein).  A
key assumption of most weak lensing analyses is that galaxies are
randomly oriented. If the shapes of the observable parts of galaxies
depend on their large-scale environment then this assumption is
violated with potentially serious consequences for inferences derived
from weak lensing methods.  The sensitivity of weak lensing to
violations of the random-orientation hypothesis is stronger if one
attempts to use lensing tomography \cite{1999ApJ...522L..21H}, as
essentially all proposed surveys wish to do.

There is agreement within the theoretical community that the {\it shapes\/}
of dark matter halos do exhibit alignment correlations, driven by the
environment in which they form.  This alignment is thought to be the
theoretical analog of the alignment correlation seen for massive elliptical
galaxies by
\cite{2002MNRAS.333..501B,Lee:2001vz,2004MNRAS.347..895H,2006MNRAS.370.1008M,2008arXiv0811.1995F}
and the models found in \cite{Heymans:2006nu} are in good agreement
with these latter measurements. These alignments seem to persist to very large
scales ($\sim$ 50 Mpc) which may indicate that the commonly adopted
approximation of Gaussian tidal fields with negligible non-linear effects
might not be fully justified \cite{2002astro.ph..5512H,2008ApJ...681..798L}.

Both the observational case for, and the theoretical models of, the
alignment of disks or angular momentum directions is less settled.
Standard assumptions are that galactic disks lie perpendicular to the
angular momentum direction of the baryons, or the inner halo, or the
entire halo.
Some hydrodynamic simulations show that that there is a strong, but not
perfect, correlation between the angular momentum vector of gas and that
of the dark matter halos hosting the galaxy \citep{2002ApJ...576...21V}.
Other simulations indicate that while the disk is very well aligned with
the halo orientation in the inner regions of the dark mater halos
(at distances less than 10\% of the virial radius), the angular momenta
between the inner and outer regions are essentially uncorrelated
\cite{2005ApJ...627..647B}.
In any case, it is reasonable to assume that the degree of alignment of
galaxies with large-scale structure is smaller than that between dark matter
halo angular momenta and the structure.  One way of searching for such a
signature is to look at the orientation of angular momentum vectors on the
outskirts of cosmic voids.
The gravitational evolution of matter fluctuations in the cosmological
context tends to make under-densities (voids) rounder, while evolving
over-densities (halos) into more elongated structures
\cite{2004MNRAS.350..517S}.
This larger degree of symmetry in the voids makes more amenable to analysis.
The theoretical situation here is also complex, though $N$-body simulations,
in general, find a very small or negligible correlation between the angular
momentum of the inner parts of dark matter halos in thick shells around voids
and the direction to the center of the void.  In particular, the
authors of references \cite{Heymans:2006nu} and \cite{Patiri:2005ys}
found no alignment.  The authors of reference \cite{Brunino:2006ym}
found an alignment only in very thin shells around the void for halos
with quiescent merging histories, not for the thicker shells used in
most observational work.
They also claim very strong numerical requirements for the method to converge.
On the other hand, the authors of \cite{Cuesta:2007it} found an alignment if
they considered the outer parts of the halo in their calculation of the
angular momentum (but not the inner regions alone), even with thick shells
and less stringent numerical requirements than \cite{Brunino:2006ym}.
Reference \cite{Hahn:2007ui} does not study the same problem, but
those authors found that halos in sheets tend to have angular momenta
parallel to the sheets, though the overall level of alignment is low.
In \cite{Paz:2008na} the authors find a range of behaviours ranging
from angular momentum pointing in the direction perpendicular to the
structure, to no-alignment, and then to a regime where the angular
momentum is contained by the plane defined by the surrounding
structure, depending on the halo mass.

Given this theoretical background, the observational situation is intriguiging.
Recently \cite{2006ApJ...640L.111T} presented a measurement from the 2dFGRS and
SDSS that spiral galaxies located on the shells of the largest cosmic voids
have rotation axes that lie preferentially in the void surface.
The strength of this alignment was much higher than all measurements of angular
momentum correlations of dark matter halos in $N$-body simulations, which as
we argued above are expected to be an upper limit to the correlations of
galaxies.
In this paper we try to update this analysis with the most recent SDSS data.





\section{Theory}

The most commonly invoked theory for the origin of galaxy spins is known
at the Tidal Torque Theory (TTT)
\citep{1969ApJ...155..393P,Dorosspin,1984ApJ...286...38W,1988MNRAS.232..339H,1996MNRAS.282..436C},
which essentially states that the dark matter spins up as a result of
coupling of the local quadrupole moment of matter distribution to the
shear field, giving
\begin{equation}
  L_i(t) = a^2(t) \dot{D} \epsilon_{ijk} T_{jl}I_{lk},
\end{equation}
where $a$ is the scale factor of the Universe, $D$ is the growth
factor and $\epsilon_{ijk}$ is the Levi-Civita symbol. The local
inertia tensor $I_{ij}$ of the protohalo (the mass that will later
form the dark matter halo) in Lagrangian space is given by
\begin{equation}
  I_{ij} = \bar{\rho_o} \int_V q_i q_j \rmd^3 q,
\end{equation}
where $q_i$ are the Lagrangian coordinates around the centre of mass
of the halo and $\rho_o$ is the mean density.  The local shear tensor
$T_{ij}$ is defined by
\begin{equation}
  T_{ij} = \partial_i \partial_j \phi (\vec{q}),
\end{equation}
where $\phi$ is the gravitational potential. 

Lee and Pen \cite{2000ApJ...532L...5L} have proposed the following simple
ansatz for the relation between the galaxy spin vector $\vec{L}$ and the
shear tensor $\vec{T}$
\begin{equation}
\left\langle \hat{L}_i \hat{L}_j \right\rangle =
  \frac{1+c}{3}- c \widehat{T}_{ik}\widehat{T}_{kj},
\label{eq:lee}
\end{equation}
where $\hat{\vec{L}}=\vec{L}/|\vec{L}|$ and $\widehat{T}$ is
defined by
\begin{equation}
  \widehat{T}_{ij} = \frac{\tilde{T}_{ij}}{\tilde{T}_{kl}\tilde{T}_{kl}},
\label{eq:rescale}
\end{equation}
with $\tilde{T}_{ij} = T_{ij}-{\rm Tr}(\vec{T}) \delta_{ij}/3$.  The
parameter $c$ which can take values between 0 and 1 describes the
effect of non-linearities. The value of $c=3/5$ corresponds to
$\vec{T}$ and $\vec{I}$ being fully uncorrelated. This ansatz is
derived from ``marginalising'' over all possible moments of inertia in
the limit that the tidal field and moment of inertia are completely
uncorrelated. The parameter $c$ was then introduced to account for the
randomizing effects of small-scale physics.

We can now build a simple toy model for the behaviour of angular
momenta in the vicinity of voids within the context of TTT and ansatz
of Equation \eqref{eq:lee}. Assuming a sharp boundary, the potential
just outside the void is 
$\phi \propto +1/r$ in comoving coordinates. Taking derivatives,
evaluating the resulting $T_{ij}$ at $(0,0,1)$ and rescaling according
to the Equation \ref{eq:rescale}, one obtains
\begin{equation}
  \widehat{T}={\rm  diag}(\,-1/\sqrt{6},\ -1/\sqrt{6},\ +2/\sqrt{6}\,).
\end{equation}
This result now holds for any $\widehat{T}$ in the frame of its
eigenvectors, with $\hat{z}$ pointing in the radial direction. Using
Equation \eqref{eq:lee} we then find
\begin{equation}
 \left\langle L_i L_j\right\rangle =  C_{ij} =  {\rm diag}
  \left(\,\frac{2+c}{6},\ \frac{2+c}{6},\ \frac{1-c}{3}\,\right). 
\end{equation}
Following the literature, we next assume that the probability
distribution function for $\hat{L}$ can be described as a Gaussian,
with zero mean and second moments $\left\langle L_i L_j\right\rangle$:
\begin{equation}
P (\hat{L}) = \frac{1}{\left(2\pi\right)^{3/2} \det {\rm C}}
\exp\left[-\frac{\hat{L} {\rm C}^{-1} \hat{L}}{2}\right].
\end{equation}

Writing $\hat{L}$ in spherical coordinates and integrating over radial
and azimuthal coordinates, one can obtain the probability distribution
function for $P(\theta)$, where $\theta$ is the remaining zenith
coordinate, namely the angle between the axis of $\hat{\vec{L}}$ and
radial direction from the centre of void:
\begin{equation}
P(\theta) =  \sin \theta\ \frac{2 (1-c) \sqrt{2+c}}{\left( 2 + 3 c \cos^2
    \theta - 2c \right)^{3/2}}.
\label{eq:1}
\end{equation}
In this equation, $P(\theta)$ is normalised so that the probability
integrates to one over the interval $\theta=0\ldots\pi/2$.  For $c=0$,
$P(\theta)=\sin \theta$, which is a purely geometrical factor. A
result applicable to a more general settings can be found in
\cite{2004ApJ...614L...1L}.

The purpose of this paper is to test whether there is any evidence
that $P(\theta)$ deviates from the simple sine form in the vicinity
of voids and if so, what values can the associated parameter $c$
take. We stress again that the voids boundaries are not sharp (as seen in
the Figure \ref{fig:voidplot2}), but fitting the Equation \eqref{eq:1}
to the data should nevertheless give some idea about the size of the
parameter $c$.

\section{Data and method} \label{sec:data-method}

Our method is essentially the same as that of
\cite{2006ApJ...640L.111T} and is composed of three steps.  First we
identify the voids in the SDSS data.  Second, we locate spiral
galaxies in the vicinity of these voids. Third, we investigate if the
distribution of spin axes of nearly face-on or edge-on galaxies is
inconsistent with random distribution. These steps are discussed in
more details in the following subsections. Throughout we use
concordant cosmology with $\Omega_\Lambda=0.75$, $\Omega_m=0.25$ and
the reduced Hubble's constant $h=H_0/100 {\rm km/s/Mpc}=0.7$.  Our
data come from the 6th data release of the SDSS
\cite{2000AJ....120.1579Y,2008ApJS..175..297A} and we use the Value
Added Galaxy Catalog (VAGC) \cite{2005AJ....129.2562B} to obtain the
volume-limited samples that are used to find voids. We limit ourselves
to the northern galactic cap.

\begin{figure}
  \centering
  \includegraphics[width=\linewidth]{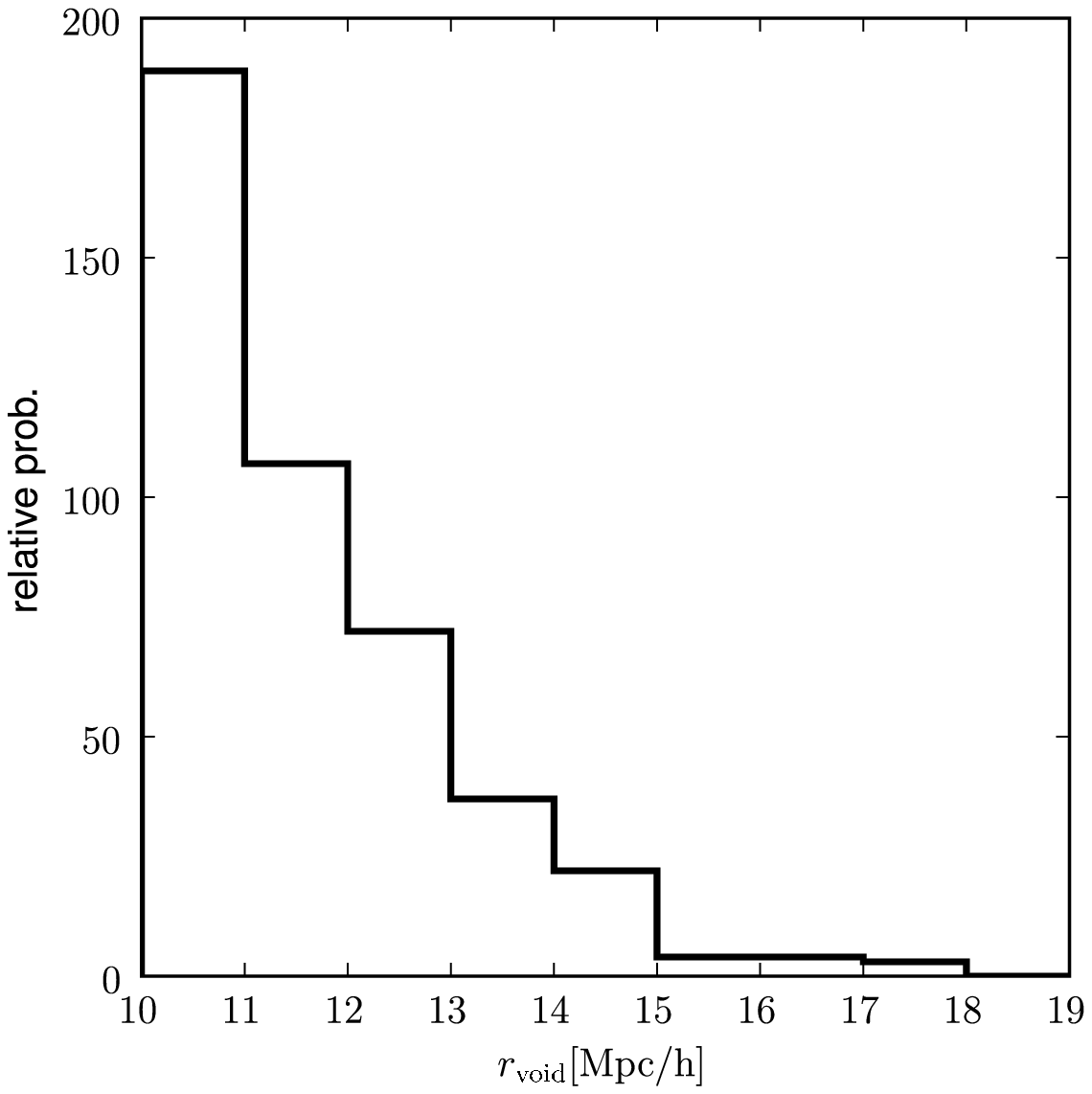}\\
\vspace*{0.4cm}
  \includegraphics[width=\linewidth]{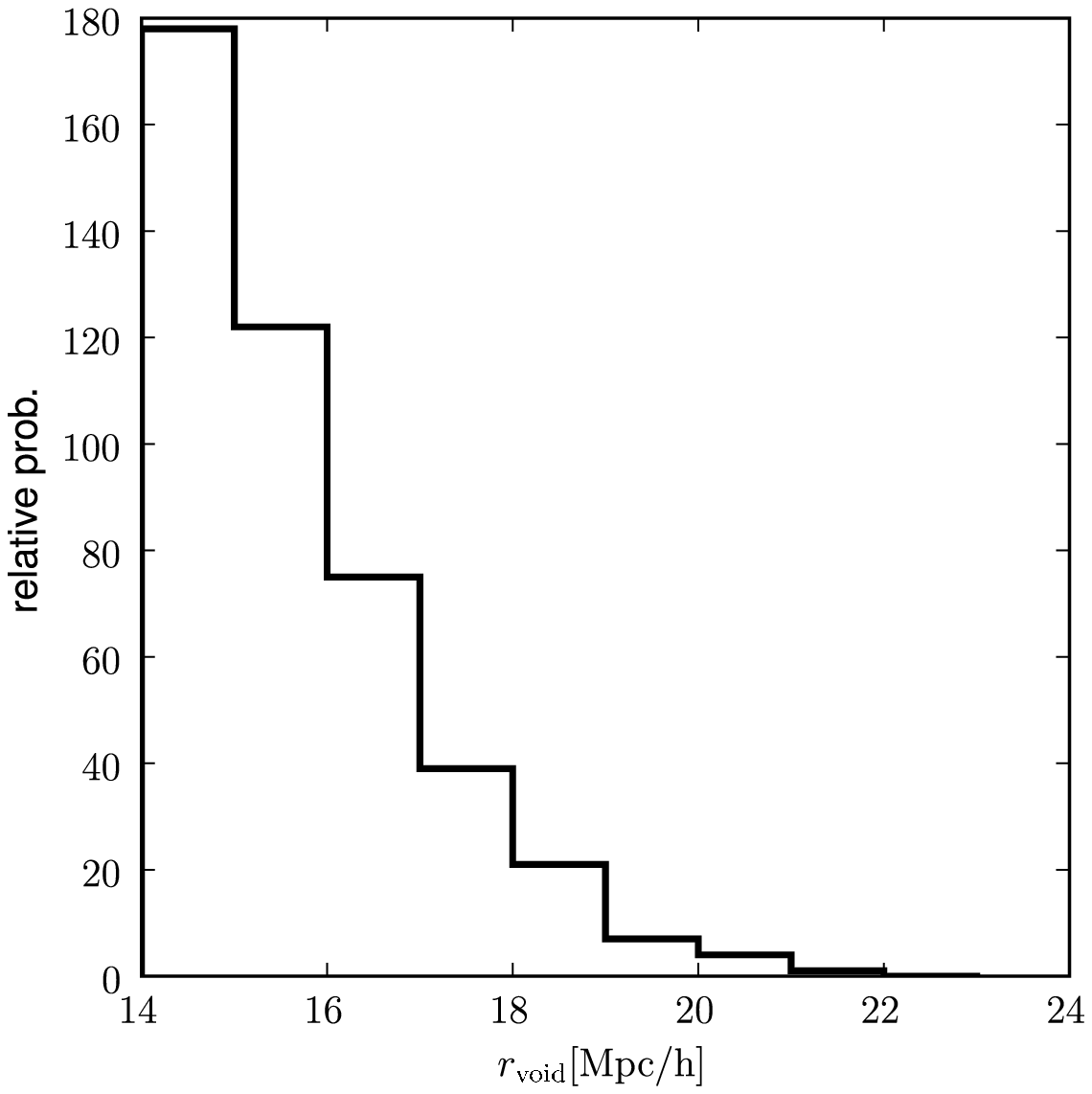}
  
  \caption{This figure shows the distribution of voids sizes for
    $m_L=-21$ (top) and $m_L=-22$ (bottom). Note that the minimum
    void size is an input parameter for the void finder and equals
    to 10 and 14 Mpc/$h$ respectively.}
  \label{fig:voidplot1}
\end{figure}

\begin{figure}
  \centering
  \includegraphics[width=\linewidth]{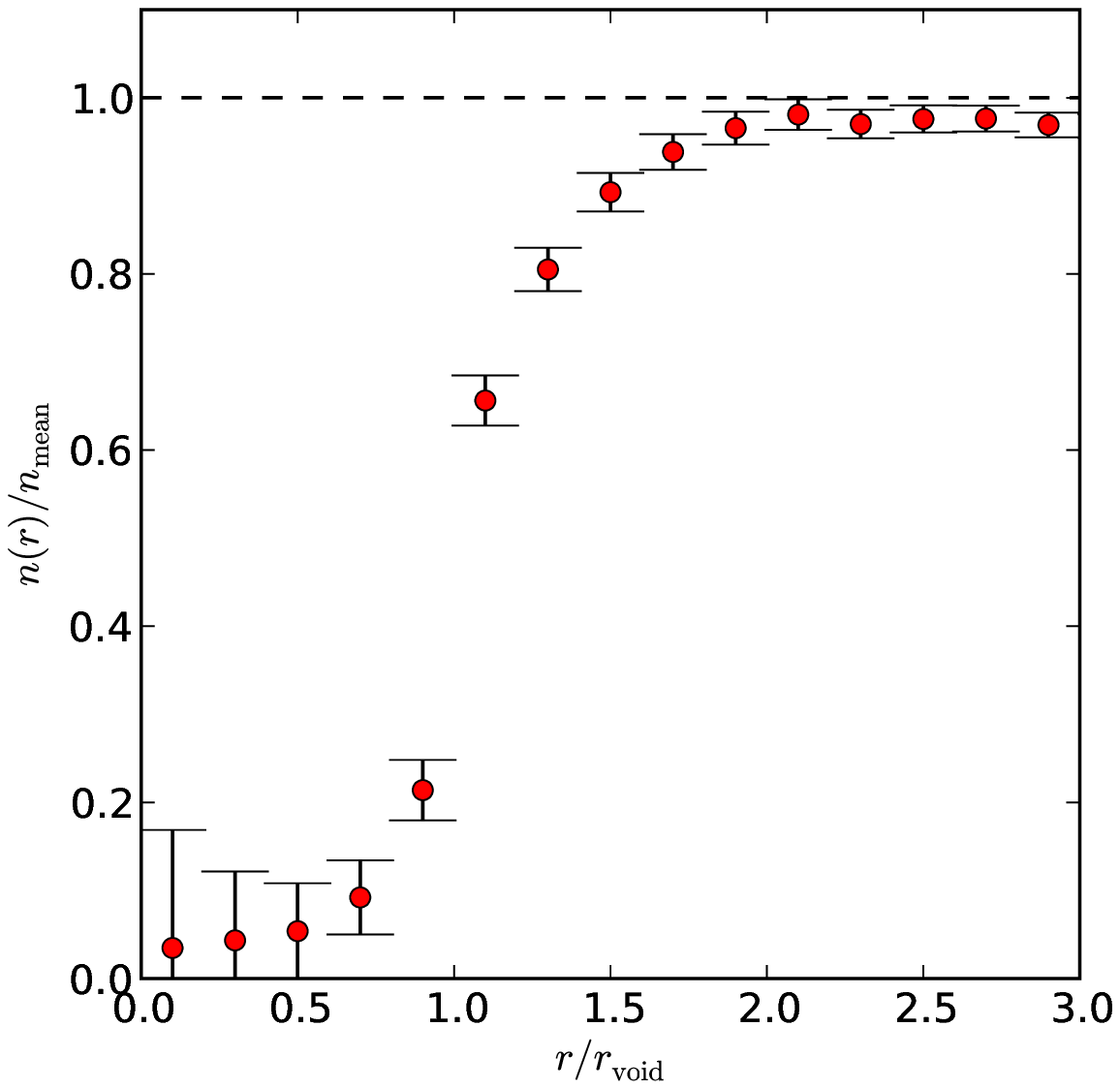} 
  \includegraphics[width=\linewidth]{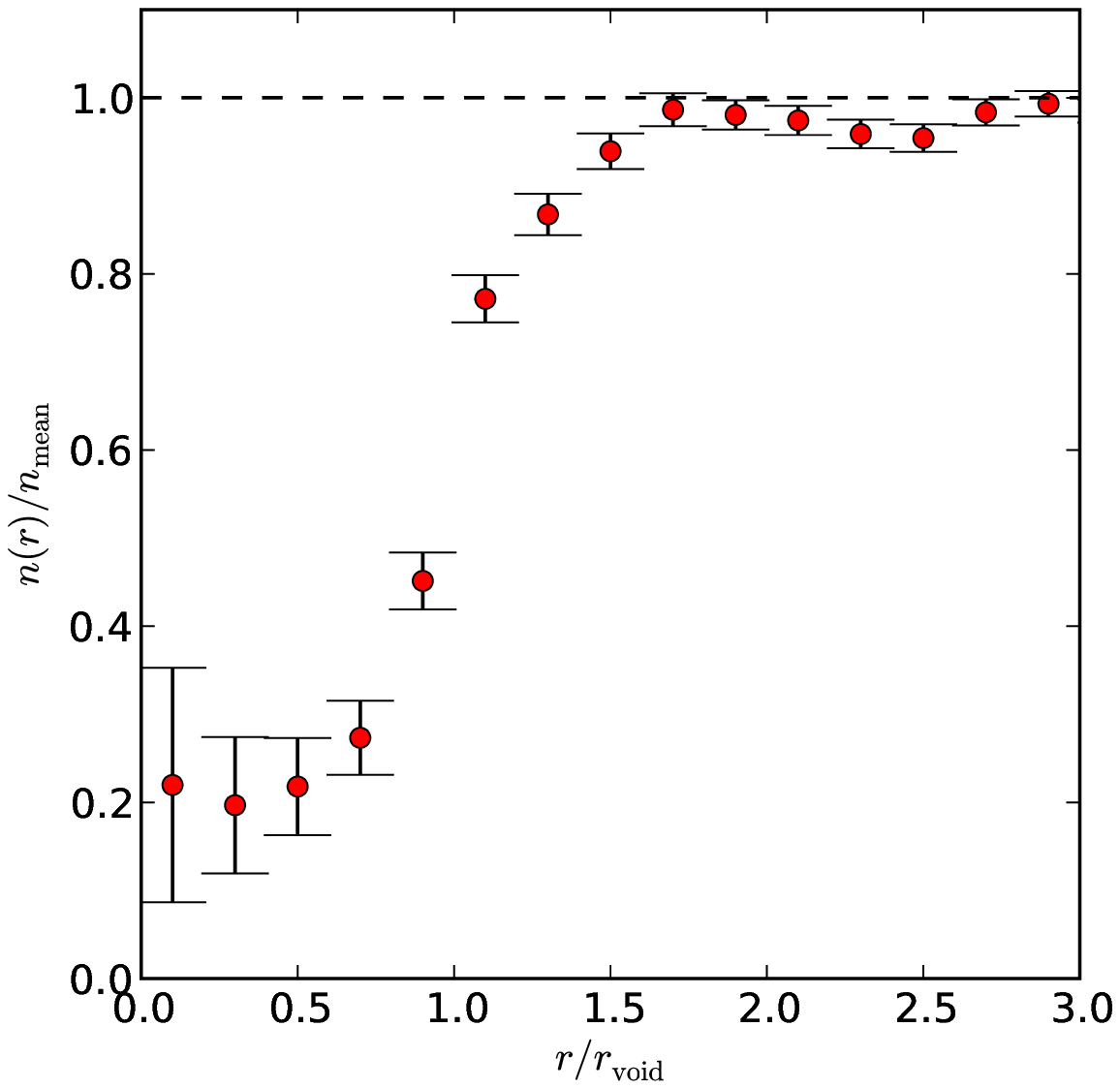}
  
  \caption{This figure shows how the mean number density of galaxies
    change as a function of distance from the void centre scaled by
    the void radius for the $m_L=-21$ (top) and $m_L=-22$ (bottom)
    samples.  Error-bars are derived from Monte Carlo simulations in
    which void centres are randomly positioned within the volume limited
    catalogue and are approximate and heavily correlated.}
  \label{fig:voidplot2}
\end{figure}

\subsection{Void Finder}

The void finder is conceptually based on the HB void finder of
\cite{Patiri:2005ys}. We start by creating a volume limited catalogue
of galaxies.  We create two volume limited catalogues, depending on
the limiting absolute extinction-corrected $r$-band Petrosian magnitude,
which we choose to be either $m_L=-22$ or $m_L=-21$.  We use only
parts of the survey where the survey completeness described by the
SDSS parameter \texttt{FGOTMAIN} is above 0.82 and randomly remove
galaxies to make the sample uniformly complete at this level.  In
addition we also use the VAGC's random catalogues to create random
catalogues with the same selection function as the main catalogues,
but with about ten times as many points. We then find voids using the
following algorithm.

\begin{enumerate}
\item Pick a random point in the survey volume. Locate four closest
  neighbours in the real catalogue. Determine the sphere that is
  defined by these four points.

\item Find if any other real galaxies are inside the sphere. If so,
  diminish the radius accordingly.

\item If the sphere's radius is less than $10$ Mpc$/h$ for the
  $m_L=-21$ or $14$ Mpc$/h$ for the $m_L=-22$, discard it and go
  back to 1.

\item Count the number of random points inside the sphere and if this
  number is more than two sigma below the number expected given the
  number density of random points, discard it and go back to
  1. Moreover, if the first moment of radial vectors of random points
  inside the void is inconsistent with random, discard them and go
  back to 1.  This step essentially ensures that the void is fully
  within the survey.

\item Compare the void candidate to the existing voids. If it touches
  any of the existing voids, then discard the smaller of the two.
  Check if the new void touches any other voids and discard those as
  well.  This ensures that two close small voids are merged into a
  bigger one if possible.
\end{enumerate}

This process is repeated for all random points. Its result is a
catalogue of non-overlapping spherical voids that contain no galaxies
that are brighter than $m_L$ (though less bright galaxies can be in the
voids).  Although not strictly ensured by the algorithm, we find that the
order in which random points are taken does not matter - the number of
void solutions that are affected by this is less than 0.5\%.
Some basic properties about the voids the we find are tabulated in the
Table \ref{tab1}.

\begin{table*}
  \begin{tabular}{c|ccc|ccc}
& & $m_L=-21$ & & & $m_L=-22$ & \\
\hline
Number of voids & & 438 & & & 447 &   \\
Minimum void radius & & 10 Mpc$/h$ & & & 14 Mpc/$h$ &  \\
Maximum void radius & & 17.8 Mpc$/h$ & & & 21.4 Mpc/$h$     &  \\
redshift range  & &  0.169 -- 0.183 & &  & 0.111 -- 0.123 &  \\
\hline
& face-on & edge-on & both & face-on & edge-on & both \\
\hline
Number of galaxies & 255 & 323 & 578 & 151 & 107 & 258 \\
Na\"{i}ve $\chi^2$ & 25.0 & 16.4 & 17.6 & 21.8 & 20.0 & 22.5 \\
MC derived $\chi^2$ & 23.9 & 14.1 & 16.8 & 20.0 & 18.5 & 21.8 \\
\hline
Constraints on $c$ & &  $c<0.11(0.13)$  & & & $c<0.16(0.19)$ &\\
  \end{tabular}
  \caption{This table shows the basic properties of voids that we find
    in our data and the resulting $\chi^2$ values for null
    model. The null model has no degrees of freedom and there are 20
    datapoints. The 95\% limits correspond to $\chi^2$ of 9.6 -- 34.2.
    The bottom section shows 95\% (99.7\%) limits on the $c$ parameter.
    See text for discussion.
  }
  \label{tab1}
\end{table*}

To aid systematic checks we also form an ensemble of random void
catalogues. In this catalogue we take the real number of voids and
their radii and distribute them non-overlappingly, but otherwise
randomly within the survey volume.

Throughout we ignore the distinction between real space and redshift
space. Since our voids are relatively large and separated from
high-density peaks, this should be a relatively safe approximation.

We plot basic properties of our voids in Figures \ref{fig:voidplot1}
and \ref{fig:voidplot2}. These are here to illustrate the properties
of our voids in broad brushes and to show the dependence of voids on
the void-finding algorithm properties. The main thrust of this paper
is to analyze the properties of galaxy spins in the vicinity of voids;
we will return to properties of voids in a forthcoming publication.

\begin{figure*}
  \centering
  \includegraphics[width=\linewidth]{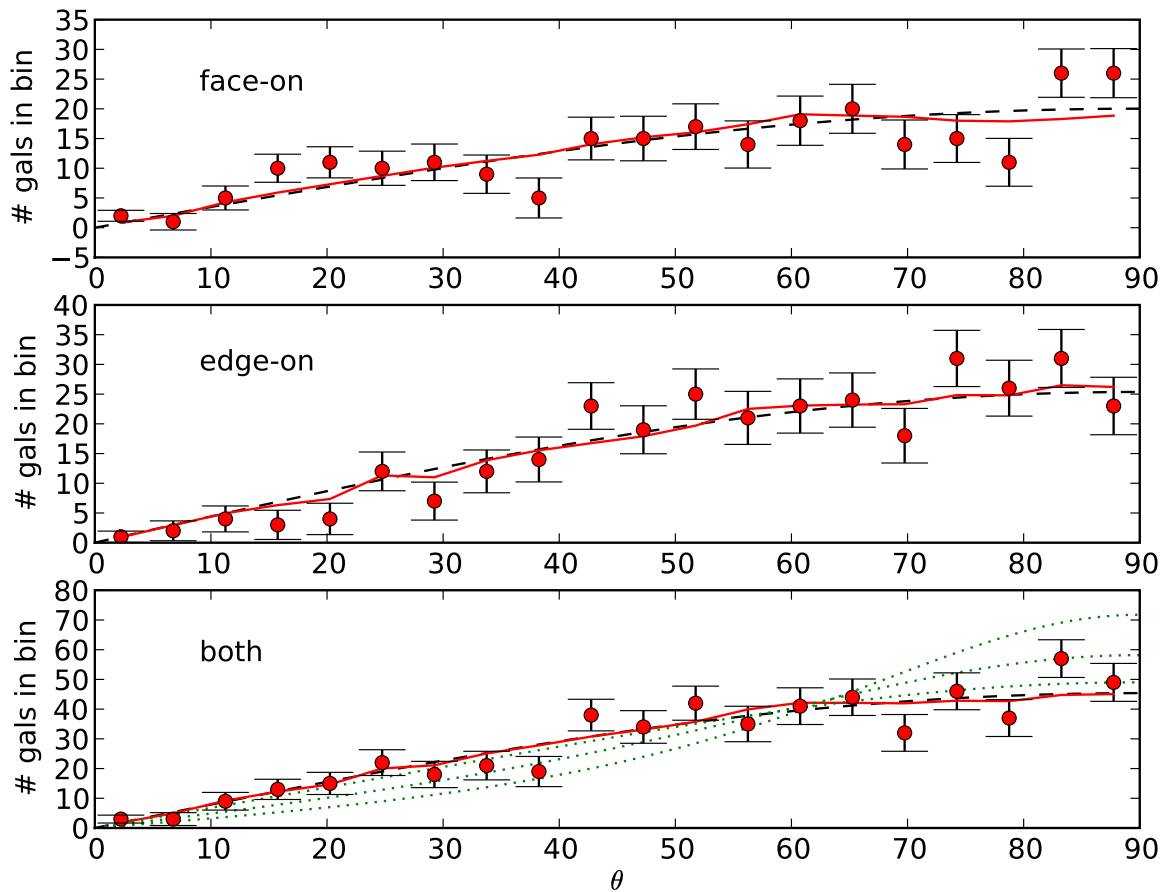}
  \caption{This figure shows the measured probability distribution for
    the angle $\theta$ in the $m_L=-21$ sample. Panels show results
    corresponding to face-on only (top), edge-on only (middle) and
    combined (bottom) spiral galaxies. The dashed black line
    corresponds to the theoretical expectation $p(\theta)=sin \theta$,
    while the solid red-line corresponds to the Monte Carlo
    results. The jaggedness in the red line is real and not a
    numerical artifact. Points are measurements with Monte Carlo
    derived errorbars. Dotted green lines are theoretical predictions
    for the models with $c=0.1,0.3,0.5$.}

  \label{fig21}
\end{figure*}

\begin{figure*}
  \centering
  \includegraphics[width=\linewidth]{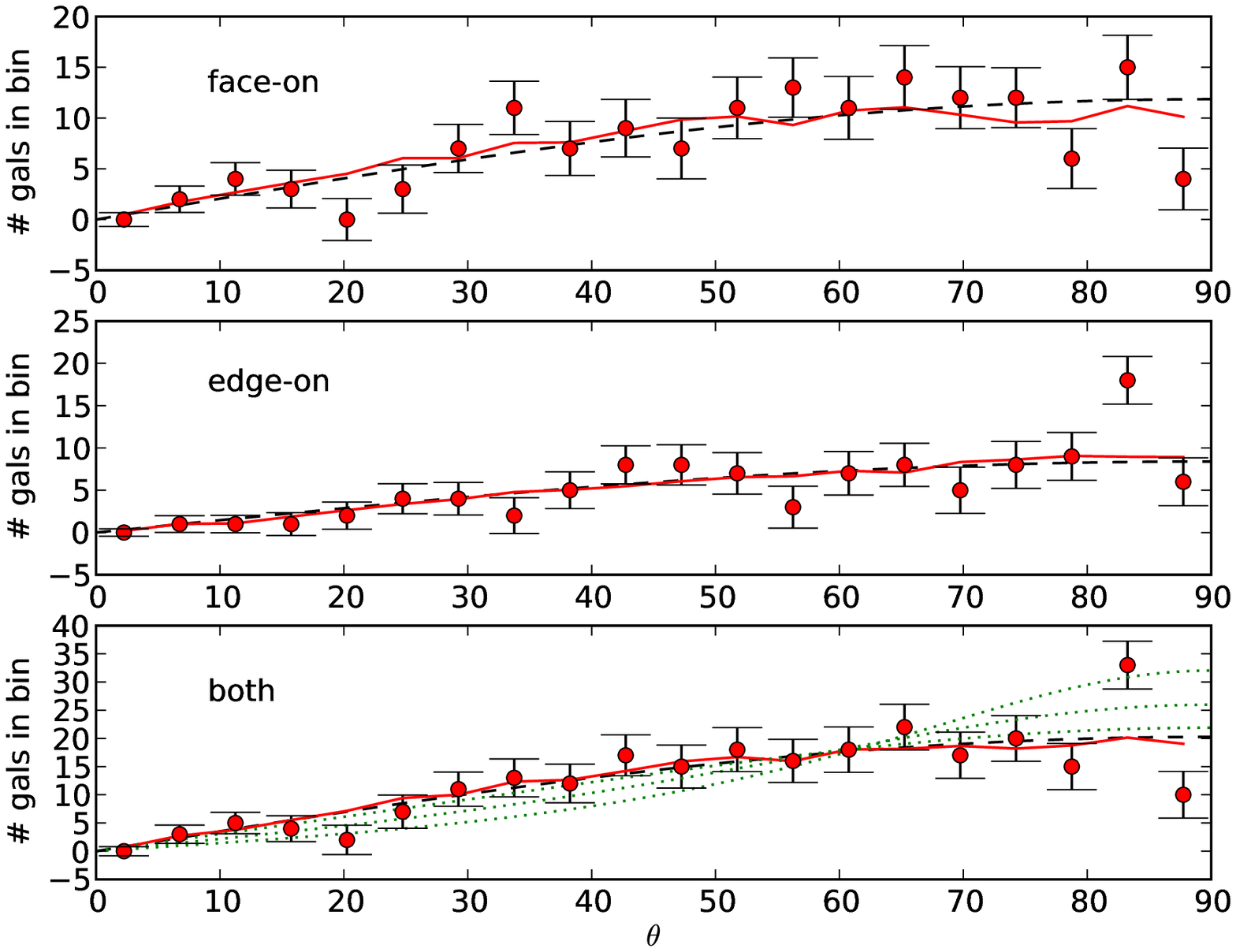}
  \caption{Same as Figure \ref{fig21} but for a catalogue of voids
    derived from $m_l=-22$ volume limited sample.}
  \label{fig22}
\end{figure*}

\subsection{Galaxies in the vicinity of voids}

We next locate spiral galaxies that reside in the vicinity of
voids. We now again work with the full SDSS catalogue and select
spiral galaxies by a colour cut $g-r<0.6$, where $g$ and $r$ are model
magnitudes from the SDSS pipeline. Spiral galaxy is considered to be
at the edge of a void if its distance to the void centre is between
$r_{\rm void}$ and $r_{\rm void}+r_{\rm taper}$, where $r_{\rm
  taper}=4{\rm Mpc}/h$ is the taper length. We limit ourselves to
either face-on or edge-on galaxies. To determine the inclination of
the galaxy, we use the adaptive moments $e_+$ and $e_\times$ from the
SDSS photometric data reduction and calculate the axis ratio $q$ using
\cite{Ryden:2003vd}:
\begin{equation}
  q = \left(\frac{1-e}{1+e} \right)^{1/2},
\end{equation}
where $e=\sqrt{e_\times^2+e_+^2}$. Galaxies whose $q>0.96$ are deemed
to be face on and those with $q<0.27$ are edge one. These criteria are
somewhat more relaxed than those of \cite{2006ApJ...640L.111T}, but
visual inspection of a random sub-sample ensured that we are indeed
selecting face-on and edge-on spiral galaxies. For face-on galaxies,
we assumed that the angular momentum axis is aligned with the radial
direction toward the galaxy. Similarly, for the edge-on galaxies it was
assumed to lie in the plane perpendicular to the radial direction and
along the galaxies apparent minor axis. The latter was again inferred
from the SDSS photometric reduction using the isophotal angle in the
$r$-band and again visually confirmed to be measuring the correct
quantity.

For the sample of galaxies and voids we then measure the probability
distribution function for $\theta$, the angle between galaxy spin and
the radial vector from the centre of the void toward the galaxy. Since
the axis of the galaxy's spin vector is a spin-2 quantity, the angle
is constrained to lie between 0$^{\circ}$ and 90$^{\circ}$. We
therefore measure the probability distribution function for this angle
in 20 bins of uniform size in this range. The distribution of $\theta$
should follow $\sin \theta$ is our sampling of galaxies around voids
would be random and if there were no edge effects associated with
voids. However, in practice, while the voids were required to lie
within the survey, the voids fattened by the taper radius might lie
outside survey. Moreover, the cosmic web makes the sampling of
positions non-uniform. In order to correct for these effects, we
Monte-Carlo (MC) our errorbars as follows. For all galaxies that lie
in the vicinity of void, including those that are neither face-on or
edge-on, we randomly permute their spin vector properties and
calculate the resulting $p(\theta)$ for an ensemble of
permutations. The question we are therefore asking is: given the
number of face-on, edge-on and remaining galaxies in the vicinity of
voids, what is the probability that this distribution is not random?
Using Monte-Carlo simulations we calculate the number of galaxies
expected in each $\theta$ bin and the correlations between adjacent
bins. Moreover, under assumption of Gaussianity, this allows us to
calculate the exact $\chi^2$ by inverting the correlation
matrix\footnote{Note that since the total number of galaxies is fixed,
  the full covariance matrix is singular. We solve this by removing
  one data point, which does not affect the $\chi^2$ value.}.

\section{Results} \label{sec:results}

Our main results are plotted in  Figures \ref{fig21} and \ref{fig22}.
These figures show the expected signal in red solid line. This was derived
using Monte Carlo simulations explained in the previous section. The
jaggedness in the line is real and corresponds to the concrete
distribution of cosmic web around the particular voids that we find. 
The plotted errorbars are also derived from the Monte Carlo Simulation.

It is clear already from the picture, that there is no strong evidence
for departure from expected signal. We quantify this in Table
\ref{tab1}.  In this table we quote two different $\chi^2$ values. The
na\"{i}ve $\chi^2$ are calculated assuming that the number of galaxies
in each bin is Poisson distributed and that bins are independent. The
Monte-carlo derived $\chi^2$s use the mean values for null theory
derived from the Monte Carlo simulations and evaluate $\chi^2$ taking
in account the full covariance properties from MC samples.  Since
there are twenty data-points at which we estimate $p(\theta)$, we note
that there are no detections at greater than $2$ sigma.

We note that the correctly calculated $\chi^2$ is in fact surprisingly
close to the na\"ively calculated one. In fact, this is due to the
large number of voids that can be found using the DR6 dataset. If we
artificially limit the number of voids, the difference can be much
larger. For example, taking the first 50 voids of the $m_L=-22$
sample, the $\chi^2$ for face-on galaxies drops from $45.3$ using
na\"{i}ve $\chi^2$ to $17.9$ using MC $\chi^2$ with
twenty degrees of freedom. This corresponds to a drop from over 3
sigma detection to less than 1 sigma. It is therefore crucial for the
spinologist's peace of mind that $\chi^2$s are calculated robustly.

Next we calculate limits on the parameter $c$, which are also shown in
the Table \ref{tab1}. Two magnitude bins give very strong, but also
very similar constraints on $c$.

\subsection{Systematics}

It is possible that there is some crucial aspect of the analysis which
is adversely affecting our results. We therefore compare the
stability of our results with respect to the following tweaks to our
data reduction and find no significant change in our results:
\begin{enumerate}

\item \emph{Including galaxies inside the void.}

\item \emph{Considering just galaxies inside the void.}

\item \emph{Changing the major to minor axes ratios required for
    face-on and edge-on criteria.} In particular, we tested using the
  same $q$ ratios as \cite{2006ApJ...640L.111T}, namely $q>0.978$ for
  face-on and $q<0.208$ for edge-on.

\item \emph{Removing the color cut}. This significantly increases our
  catalogue of galaxies, which becomes heavily contaminated by
  ellipticals.
\end{enumerate}

\section{Discussion \& Conclusion}
\label{sec:conclusions}

In this paper we have reconsidered the alignment of galaxy spin axes
in the vicinity of voids. We find no statistical departure from random
for the orientation of galaxy spin axis in the vicinity of
voids. Expressed in terms of the $c$ parameter, we find very stringent
limit on $c<0.11\ (0.13)$ at 95\% (99.7\%) confidence limits. These
limits assume a toy model with perfectly spherical voids and sharp
edges. These are only approximately true in reality, but the numbers
that we report should give an idea about the size of allowed values of
$c$.  Our limits are consistent with most $N$-body simulations, but there
is some tension with the largest signal found by \cite{Cuesta:2007it}, which
is $c\sim0.15$.

Our results are also inconsistent with previous results of
\cite{2006ApJ...640L.111T}, who find $c=0.7^{+0.1}_{-0.2}$.  The reasons
for this discrepancy are not immediately clear, but we note that the
catalogue that we use is considerably larger and has a much better
filling factor that dramatically increases the number of voids. It
possible that the results of \cite{2006ApJ...640L.111T} were simply a
statistical fluctuation.

When compared to the $N$-body simulations, the tidal torque theory seems to
give qualitatively, but not quantitatively correct results
\citep*{Porciani:2001db,Porciani:2001er}. The ansatz of Equation \eqref{eq:lee}
is perhaps less universally applicable.  It was found to work reasonably well
in early data \cite{2000ApJ...543L.107P}; however, there was recently a
tentative detection of correlation of spin vectors that should vanish in this
theory \cite{2008arXiv0809.0717S}.
In this paper we find $c$ is consistent with zero.
Moreover, there does not seem to be good evidence for decoupling of scales
between the tidal field and the local moment of inertia; a simplistic
$k$-counting gives the same scale dependence.
Finally, it is not clear whether the parameter $c$ should have one universal
value or be an environment or scale-dependent variable.

On the basis of our revisiting the alignment of galaxy spins in the vicinity
of voids we are again able to reconcile results from $N$-body simulations with
observations.
This is especially important in the light of the forth-coming weak lensing
surveys; a nagging discrepancy between simulations and observations might lower
our trust in corrections and systematic errors due to intrinsic alignment that
are derived from $N$-body simulations.
While more work must be done, the outlook is currently quite positive.

\vspace*{2cm}

\section*{Acknowledgements}

AS acknowledges funding from Berkeley Center for Cosmological Physics.


\bibliographystyle{arxiv}
\bibliography{cosmo,cosmo_preprints}

\end{document}